\documentstyle[12pt]{article}
\newcommand{\text}[1]{\mbox{\scriptsize #1}}
\newcommand{\bm}[1]{\mbox{\boldmath$#1$}}
\title{Discussion on spin-flip synchrotron radiation}
\author{V.A.~Bordovitsyn~\thanks{E-mail: astrodep@niipmm.tsu.ru}, 
V.S.~Gushchina and A.N.~Myagkii~\thanks{E-mail: myagkii@mail.ru}\\
{\it Tomsk State University, Tomsk, Russian Federation}}
\date{}
\begin{document}
\maketitle

\begin{abstract}

Quantum spin-flip transitions are of great
importance in the synchrotron radiation (SR) theory. For better understanding
of the nature of this phenomenon, it is necessary to except the effects
connected with the electric charge radiation from observation.
This fact explains the suggested choice of the spin-flip radiation model
in the form of radiation of the electric neutral Dirac-Pauli particle moving
in the homogeneous magnetic field. It is known that in this case, the total
radiation in the quantum theory is conditioned by spin-flip transitions.
The idea is that spin-flip radiation is represented as a nonstationary
process connected with spin precession. From this point of view, we shall
shown how to construct a solution of the classical equation of spin
precession in the BMT theory having the exact solution of the
Dirac-Pauli equation. Thus, one will find the connection of
the quantum spin-flip transitions with classical spin precession.

\end{abstract}

According to the uncertainty principle, for the spin-flip transition with
the characteristic frequency of SR~\cite{1} we obtain
$$\omega_{\max}\approx\omega_0\gamma^3 =\frac{2\vert\mu\vert H}{\hbar}\;
\gamma^2,\quad \omega_0=\frac{eH}{m_0 c\gamma}\approx\frac{c}{\rho},$$
where $\mu$ is the magnetic moment, $\rho$ is the radius of curvature of
the relativistic electron trajectory in the homogeneous magnetic field $H$.
The ratio of the transition time $\Delta\tilde t$ and the time of the
radiation forming $\Delta t$ on the circular arc $dl\approx\gamma\rho$
is equal
$$\Delta\tilde t/\Delta t\approx 2\pi\gamma^{-4} \ll 1.$$

That means that in the ultrarelativistic case, spin-flip transitions
are practically non-inertial, that is why the electron motion can be
considered for the time $\Delta\tilde t$ as uniform and rectilinear.

Let us consider the electric neutral Dirac-Pauli particle,
which moves in the homogeneous magnetic field ${\bm H}=(0,0,H)$.
As is known~\cite{2}, wave function of the above particle has the form
$$\Psi({\bm r},t)=L^{-3/2}\left(
\begin{array}{l}
c_1\\
c_2e^{i\varphi}\\
c_3\\
c_4e^{-i\varphi}
\end{array}
\right) \exp(ib_\mu\xi^\mu)=\psi_\zeta ({\bm r})
\exp\left(-i\frac{m_0c^2}{\hbar}\gamma_\zeta t\right),$$
where $b^\mu =\gamma_\zeta (1,{\bm\beta})$, $\xi^\mu =(m_0 c/\hbar)r^\mu$
are the dimensionless energy-momentum and the coordinates of the particle,
$$\gamma_\zeta =\gamma \left( 1+\zeta S\sqrt{1+\gamma^2\beta_\perp^2}\right)
,\quad \zeta=\pm 1,$$
and moreover, $S=\vert\mu\vert H/m_0 c^2$, $\gamma =1/\sqrt{1-\beta^2}$.

The spin coefficients $c_i\ (i=1,2,3,4)$ are determined from the condition
that the stationary wave function $\psi_\zeta$ is an eigenfunction of the
spin operator~\cite{2}
$$\widehat\Pi_z =\rho_2 [{\bm\sigma}\widehat{\bm b}]_z +\sigma_z ,$$
which is a motion integral and describes the spin projection on the
magnetic field direction with the quantum number $\zeta$.

If the particle moves in the plane $XZ\ (\varphi =0)$, then for
${\bm\beta}= \beta (\sin\alpha ,0, \cos\alpha)$ the spin
coefficients take on the form
$$ \left( \begin{array}{c} c_1\\ c_2\\ c_3\\ c_4 \end{array}
\right)=\left(
\begin{array}{r}
\frac{\zeta}{2}\sqrt{\frac{1}{2}\left( 1+\frac{\zeta}{q}\right) }
\left(\sqrt{1+\beta_z}+\zeta\sqrt{1-\beta_z}\right) \\
-\frac{1}{2}\sqrt{\frac{1}{2}\left( 1-\frac{\zeta}{q}\right) }
\left(\sqrt{1+\beta_z}-\zeta\sqrt{1-\beta_z}\right) \\
\frac{\zeta}{2}\sqrt{\frac{1}{2}\left( 1+\frac{\zeta}{q}\right) }
\left(\sqrt{1+\beta_z}-\zeta\sqrt{1-\beta_z}\right) \\
\frac{1}{2}\sqrt{\frac{1}{2}\left( 1-\frac{\zeta}{q}\right) }
\left(\sqrt{1+\beta_z}+\zeta\sqrt{1-\beta_z}\right)
\end{array}
\right) =\vert\zeta \rangle,$$
where $q=\gamma\sqrt{1-\beta^2\cos^2\alpha}$. We note that the spin
coefficients written here in the first approximation on $\hbar$ differ
from zero.

In the quantum theory the spin precession is described by the nonstationary
wave function which is a superposition of the spin states $\zeta =\pm 1$
$$\widetilde\Psi ({\bm r},t)=A\psi_1({\bm r})\exp\left\{-i\frac{m_0c^2}{\hbar}
\gamma_1 t\right\}+
B\psi_{-1}({\bm r})\exp\left\{-i\frac{m_0c^2}{\hbar}\gamma_{-1}
t\right\}.$$

The constant factors $A$ and $B$ are determined from the initial condition
$$(\widehat{\bm\Pi}{\bm n})\widetilde\Psi ({\bm r},0)=\lambda\widetilde\Psi
({\bm r},0),$$
where ${\bm n}=(\sin\lambda\cos\nu ,\sin\lambda\sin\nu ,\cos\lambda)$.
For example, in the case of the initial spin orientation along the $X$-axis,
we shall have
\begin{eqnarray*}
A& = &-\frac{\varepsilon}{\sqrt{2}}\;\sqrt{1-\frac{\varepsilon\gamma\beta^2
\sin\alpha\cos\alpha}{\sqrt{1+\gamma^2\beta^4\sin^2\alpha\cos^2\alpha}}},\\
B& = &\frac{1}{\sqrt{2}}\;\sqrt{1+\frac{\varepsilon\gamma\beta^2
\sin\alpha\cos\alpha}{\sqrt{1+\gamma^2\beta^4\sin^2\alpha\cos^2\alpha}}},
\end{eqnarray*}
$$\lambda =\varepsilon\sqrt{1+\gamma^2\beta^2\cos^2\alpha},$$
along the $Y$-axis
$$A=\frac{\varepsilon}{\sqrt{2}},\quad B=\frac{-i}{\sqrt{2}},\quad
\lambda=\varepsilon\gamma,$$
along the $Z$-axis
$$A=(1,0),\quad B=(0,1),\quad \lambda =\zeta\;\frac{1+\gamma^2\beta^2
\sin^2\alpha}{\gamma\sqrt{1-\beta^2\cos^2\alpha}}.$$

We note that in the second case, these relations have the simplest form, as
the spin tensor component $\widehat{\Pi}_y$ is really not subjected to the
Lorentz transformations. Value $\varepsilon=\pm 1$
characterizes the direction of the initial spin orientation. In the
latter case, $\varepsilon =\zeta$.

To calculate the mean value of all components of $\widehat{\Pi}{}^{\mu\nu}$
tensor it is also necessary to use the relations
\begin{eqnarray*}
\langle\zeta\vert\widehat{\Pi}_x\vert\zeta \rangle=
-\zeta\frac{\gamma\beta^2\sin\alpha\cos\alpha}
{\sqrt{1-\beta^2\cos^2\alpha}},&&
\langle-\zeta\vert\widehat{\Pi}_x\vert\zeta
\rangle=-\frac{1}{\sqrt{1-\beta^2\cos^2\alpha}},\\
\langle\zeta\vert\widehat{\Pi}_y\vert\zeta \rangle=0,&&
\langle-\zeta\vert\widehat{\Pi}_y\vert\zeta \rangle=-i\zeta\gamma,\\
\langle\zeta\vert\widehat{\Pi}_z\vert\zeta
\rangle=\zeta\gamma\sqrt{1-\beta^2\cos^2\alpha},&&
\langle-\zeta\vert\widehat{\Pi}_z\vert\zeta \rangle=0.
\end{eqnarray*}

The mean values of the components $\langle\widehat{\bm\Pi}\rangle_t$
will essentially differ from one another in dependence on the initial spin
orientation. In the most common case when
$\langle\widehat{\bm\Pi}\rangle_0=
\langle 0,\widehat{\Pi}_y,0\rangle_0$ we have
$$\langle\widehat{\Pi}_x\rangle_t=-\frac{\varepsilon}
{\sqrt{1-\beta^2\cos^2\alpha}}\sin\omega t,$$
$$\langle\widehat{\Pi}_y\rangle_t=\varepsilon\gamma\cos\omega t,\quad
\langle\widehat{\Pi}_z\rangle_t=0.$$

As a matter of fact these expressions are not different from the solution
of the tensor BMT-equation of spin precession~\cite{3}.

The frequency of spin precession determined by the condition
$$\omega =\frac{m_0c^2}{\hbar}\;(\gamma_\zeta -\gamma_{-\zeta})=\zeta\;
\frac{2\vert\mu\vert H}{\hbar}\;\sqrt{1-\beta^2\cos^2\alpha}.$$

For the initial spin orientation along the $X$ axis all calculations omitted
here become more complex although nothing changes in principle.

Interesting results are obtained at the initial spin orientation along the
$Z$ axis
$$\langle\widehat{\Pi}_x\rangle_t=-\zeta\;
\frac{\gamma\beta^2\sin\alpha\cos\alpha}{\sqrt{1-\beta^2\cos^2\alpha}},$$
$$\langle\widehat{\Pi}_y\rangle_t=0,\quad
\langle\widehat{\Pi}_z\rangle_t=\zeta\gamma\sqrt{1-\beta^2\cos^2\alpha},$$
that is all the components are constant.

Calculation of the spin invariant is of great interest
$$I=\frac{1}{2}\langle\widehat{\Pi}_{\mu\nu}\rangle_t
\langle\widehat{\Pi}^{\mu\nu}\rangle_t=
\frac{1}{\gamma^2}
\langle\widehat{\bm\Pi}\rangle_t^2+
\langle(\bm\beta\widehat{\bm\Pi})\rangle_t^2.$$

Simple but unwieldy calculations show that in any case $I=1$, which
corresponds to the classical expression of this invariant.

All results obtained are in complete agreement with the classical
theory of the spin precession on the base of the BMT equation~\cite{3}.
This method allows to determine the precession of the spin
projection on the direction of motion
$\langle({\bm\beta\widehat{\bm\Pi}})\rangle_t$.
Again all results coincide with the classical theory
$$\langle({\bm\beta\widehat{\bm\Pi}})\rangle_t=\frac{\varepsilon\beta}
{\gamma^2(1-\beta^2\cos^2\alpha)}\;
(\cos^2\alpha +\gamma^2\sin^2\alpha\cos\omega t).$$

What do the results obtained here to do with the spin-flip radiation
theory? It is known~\cite{1,4} that in the classical theory spin or,
more precisely, the intrinsic magnetic moment, orientated towards or opposite
the field direction does not radiate during uniform and rectilinear particle
motion. This radiation is accounted for by the spin precession in the plane
orthogonal to the magnetic field and all radiation characteristics
completely coincide with the quantum theory of spin-flip
radiation~\cite{2}. It follows from the above that, in fact, there is no
difference between classical and quantum interpretation of the radiation
in the nonstationary representation of spin-flip transitions.

\end{document}